\pdfoutput=1

\documentclass[11pt]{article}

\usepackage[preprint]{acl}

\usepackage{times}
\usepackage{latexsym}

\usepackage[T1]{fontenc}

\usepackage[utf8]{inputenc}

\usepackage{microtype}

\usepackage{inconsolata}

\usepackage{graphicx}
\usepackage{amssymb}
\usepackage{amsmath} 
\usepackage{graphicx}
\usepackage{mathtools}  
\usepackage{cleveref}
\Crefname{figure}{Fig.}{Figs.}
\usepackage{booktabs}
\usepackage[linesnumbered,ruled,vlined]{algorithm2e}
\SetKwInput{KwInput}{Input}
\SetKwInput{KwOutput}{Output}
\SetKw{KwEach}{for each}
\SetAlgoNlRelativeSize{-1}

\title{Structure-Aware Corpus Construction and User-Perception-Aligned Metrics for Large-Language-Model Code Completion}

\author{
 \textbf{Dengfeng Liu\textsuperscript{1}\footnotemark[1]},
 \textbf{Jucai Zhai\textsuperscript{1}\footnotemark[1]\footnotemark[2]},
 \textbf{Xiaoguang Jiang\textsuperscript{1}},
 \textbf{Ziqun Li\textsuperscript{1}},
 \textbf{Qianjin Yu\textsuperscript{1}},
\\
 \textbf{Feng Liu\textsuperscript{1}},
 \textbf{Rui Ye\textsuperscript{1}},
 \textbf{Huang Liu\textsuperscript{1}},
 \textbf{Zhiguo Yang\textsuperscript{1}},
 \textbf{Yongsheng Du\textsuperscript{1}},
 \textbf{Fang Tan\textsuperscript{1}},
\\
\\
 \textsuperscript{1}
Intelligent System Department, Zhongxing Telecom Equipment(ZTE), Changsha, Hunan, China
\\
 \small{
   \textbf{Correspondence:} \href{mailto:email@domain}{jucaizhai@gmail.com}, \href{mailto:email@domain} {liudengfeng9912@gmail.com}
 }
}

\begin{document}
\maketitle

\renewcommand{\thefootnote}{\fnsymbol{footnote}} 
\footnotetext[1]{These authors contributed equally to this work.} 
\footnotetext[2]{Corresponding author.} 

\begin{abstract}
Code completion technology based on large language model has significantly improved the development efficiency of programmers. However, in practical applications, there remains a gap between current commonly used code completion evaluation metrics and users' actual perception. To address this issue, we propose two evaluation metrics for code completion tasks—LCP and ROUGE-LCP, from the perspective of probabilistic modeling.
Furthermore, to tackle the lack of effective structural semantic modeling and cross-module dependency information in LLMs for repository-level code completion scenarios, we propose a data processing method based on a Structure-Preserving and Semantically-Reordered Code Graph (SPSR-Graph). Through theoretical analysis and experimental validation, we demonstrate the superiority of the proposed evaluation metrics in terms of user perception consistency, as well as the effectiveness of the data processing method in enhancing model performance.
\end{abstract}

\section{Introduction}

In recent years, the capabilities of large language model (LLM) in code understanding and generation have made remarkable progress, greatly driving transformations in the software development field. Code assistants such as GitHub Copilot \cite{GitHubCopilot} and Cursor \cite{Cursor} have enabled efficient handling of "on-the-fly completion" tasks like single-line and in-line code completions, significantly boosting development efficiency \cite{Takerngsaksiri2024on-the-fly,qi2024tokenpredictionsufficientgpt_on_the_fly2}.

Currently, LLMs perform well in code completion task with contextual information. Despite their excellent performance on synthetic datasets, only a few models have been successfully applied to real-world products. In our practical application of ZTE-Code-Copilot (a private-domain code assistant for communication), we encountered similar issues, primarily attributed to the following two factors:

1. A discrepancy between model capabilities and users' actual perception, reflecting the limitations of current evaluation metrics in capturing user experience \cite{liu2024stall+_codecompletion1,wu2024repomastereval_codecompletion2,van2024investigating_codecompletion3};

2. Existing LLMs typically rely on simple token sequence-based training methods, focusing on local associations between tokens, while failing to effectively capture the intrinsic structure and cross-file, cross-module dependencies of code \cite{prompt,GraphCoder}.

These issues significantly limit the practical application value of code models in complex industrial environments.

To address the first issue, some researches have attempted to screen suitable evaluation metrics by incorporating user perception data and explore the relationship between evaluation metrics and user behavior \cite{Aye2021-online-eval2,Bibaev2022-online-evaluate,vandam2023-offline}. However, in practical scenarios, the design of evaluation metrics needs to balance LLM characteristics with user perception to better guide model training. Therefore, how to design effective evaluation metrics has become a key challenge in evaluating LLMs' code completion capabilities.

Inspired by this, we propose two evaluation metrics for "on-the-fly completion" tasks: LCP (Longest Common Prefix) and ROUGE-LCP. We first theoretically analyze the relationship between these metrics and the probability distributions of model outputs; subsequently, based on ZTE-Code-Copilot's data logging system, we conduct qualitative and quantitative analyses of over ten thousand user completion behaviors within two months, validating the consistency between the proposed metrics and user adoption behavior.

Regarding the second issue, existing researches enhance models' structural awareness by constructing training corpora using abstract syntax trees (ASTs) \cite{AST-T5, jiang2025aixcoder7b_em_lightweighteffectivelarge}. However, these methods are limited to local structures within single files, lacking effective perception of global cross-file dependencies. Other researches adopt Retrieval-Augmented Generation (RAG) methods to explicitly retrieve dependent code snippets for target code fragments, enhancing the model's completion capabilities \cite{GraphCoder}, but fail to fully consider context and model affinity. How to perform repository-level code structure modeling and enhance models' perception of cross-file, cross-module code structures and semantic dependencies has become an urgent problem to be solved.

To this end, we propose a repository-level code corpus processing framework. First, we design an AST-based semantic unit segmentation method, enhancing the semantic structural integrity of pre-training corpora through structural consistency checks and multi-scale granularity control. Subsequently, we construct a Structure-Preserving and Semantically-Reordered Code Graph (SPSR-Graph), explicitly modeling function-level call relationships and struct reference paths across files, thereby significantly improving the model's understanding of repository-level code. Experimental results show that our proposed method performs exceptionally well in the communication domain's "on-the-fly completion" task, significantly improving the quality and generalization ability of code completion in cross-file, cross-module scenarios.

In summary, the main contributions of this paper are as follows:

\begin{itemize}

\item We propose two evaluation metrics for "on-the-fly completion" tasks—LCP and ROUGE-LCP. Through theoretical derivation and empirical analysis, we demonstrate the consistency of the proposed metrics with user perception.

\item We propose a data processing method based on the Structure-Preserving and Semantically-Reordered Code Graph for constructing model training corpora, significantly enhancing the model's perception and utilization of cross-file, cross-module code structures and semantic dependencies. The effectiveness and superiority of the method are validated in the communication domain's "on-the-fly completion" task.
\end{itemize}

\section{Related Work}
\subsection{Evaluation Metrics for Code Completion Tasks}

Currently, many evaluation metrics are used for code completion tasks. Common metrics include Exact Match (EM), Unit Testing (UT), Edit Similarity (ES), BLEU, ROUGE-L, and ROUGE-N \cite{zhuo2024bigcodebench_ut,ding2023crosscodeeval_em_es,wang2024rlcoder_em,Izadi2024Code4me_rouge_L,yang2024evaluatingaligningcodellmshuman_codearena,austin2021MBPP_programsynthesislargelanguage}. Although unit testing is considered the gold standard for code tasks, its high production cost makes it impractical for large-scale use. For code completion tasks, researchers tend to use EM and ROUGE-L due to the availability of context and ground truth \cite{li2025aixcoder7bv2_em_trainingllmsfully,jiang2025aixcoder7b_em_lightweighteffectivelarge,Izadi2024Code4me_rouge_L}.
However, in our practice, we find that EM and ROUGE-L still fall short in capturing user perception for "on-the-fly completion" tasks.

\subsection{Repository-Level Code Completion}

Repository-level code completion scenarios often require long-context inputs spanning multiple files and modules, inevitably introducing redundant information and reducing generation quality \cite{prompt}. To address this, CoCoMIC \cite{CoCoMIC} and RepoFusion \cite{RepoFusion} enhance models' context-awareness through long-context fine-tuning of LLMs. To further reduce contextual redundancy, CodeT5 \cite{CodeT5} and PLBART \cite{PLBART} construct hard-structured code corpora using abstract syntax trees (ASTs); AST-T5 \cite{AST-T5} further constructs soft-structured code corpora with high model affinity using ASTs, improving the model's perception capabilities. GraphCoder \cite{GraphCoder} introduces control flow graphs to build Code Context Graphs and injects context through Retrieval-Augmented Generation (RAG), improving the model's accuracy in repository-level code completion tasks.

However, existing models still suffer from insufficient structural awareness when handling long-context inputs spanning multiple files and modules.

\section{Method}
\subsection{User Behavior Analysis}

\textbf{Definition of Adoption}: Developers accept the code suggestions provided by Copilot and directly apply them to the project, either as-is or after modification. Here, we consider each press of the "Tab" key by the user as one adoption.

\textbf{Definition of Adoption Rate}: (Number of adopted code suggestions / Total number of code suggestions generated by Copilot during development) * 100\%

\textbf{User Behavior Analysis}: From a practical perspective, in code completion scenarios, the higher the similarity between the model's output and the reference answer, the more likely users are to adopt it. However, considering the left-to-right code editing habits of users, they often pay special attention to whether the model's output starts correctly from the first character. Even if only part of the output is correct, users may still adopt it and manually modify the result.

Therefore, we can reasonably infer that the longer the longest common prefix (LCP) between the model's output and the reference answer, the higher the user adoption rate.

This observation suggests that traditional metrics such as Exact Match (\textsc{EM}) or \textsc{ROUGE-L} may not fully reflect the model's performance in real-world usage scenarios. Instead, metrics focusing on continuous prefix matching better align with user expectations and interaction patterns.

To address this issue, we propose two new evaluation metrics that better align with user perception:

- Longest Common Prefix (LCP);

- A variant of ROUGE-L —— ROUGE-LCP.

Next, we will analyze the theoretical foundation of these proposed metrics based on the probabilistic modeling mechanism of large language models (LLMs) and establish their probability distribution models to explain the relationship between these metrics and user adoption behavior.

\begin{figure*}[htbp]
    \centering
    \includegraphics[width=\textwidth]{ 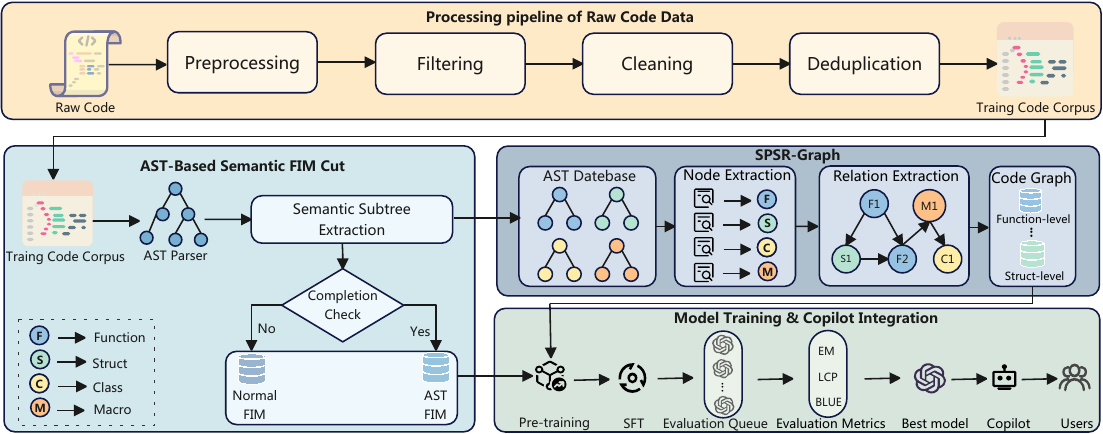}
    \caption{Overview of the proposed framework. The framework includes three stages: corpus preprocessing, AST structural segmentation, and structure-aware graph construction, transitioning from local semantic preservation to global dependency modeling to systematically construct structure-aware code completion corpora.}
    \label{fig:overall}
\end{figure*}

\subsection{Design of LCP and ROUGE-LCP Metrics}
\noindent\textbf{Longest Common Prefix (LCP)}:
The Longest Common Prefix (LCP(S, R)) is defined as the maximum number of consecutive characters that match between the model's output $S$ and the reference text $R$, starting from the beginning.

This metric emphasizes continuity starting from the first character, which is a critical feature in interactive code writing and editing. Unlike LCS-based metrics that allow non-contiguous matches, LCP aligns more closely with real user experiences.

Based on the probability formula for LLM outputs, we can model the correctness probability of the first $n$ characters of the model's output, i.e., derive the probability distribution of LCP.

Let $n$ represent the number of consecutive correct characters before the first incorrect position. Let the model's output sequence be $S = s_1, s_2, ..., s_T$, and the reference sequence (i.e., the actual code) be $R = r_1, r_2, ..., r_T$.
In other words, $n = k$ means the first $k$ characters are correct (i.e., $s_1 = r_1, ..., s_k = r_k$), while the $(k+1)$-th character is the first error (i.e., $s_{k+1} \neq r_{k+1}$).

The probability of $n = k$ is modeled as:
\begin{small}
\begin{equation*}
\begin{split}
P&(n = k) =P(LCP(S,R)=k)= \\
         &\left( \prod_{t=1}^{k} P(s_t = r_t \mid s_1 = r_1, \dots, s_{t-1} = r_{t-1}) \right) \\
         &\quad \cdot \left( 1 - P(s_{k+1} = r_{k+1} \mid s_1 = r_1, \dots, s_k = r_k) \right)
\end{split}%
\end{equation*}
\end{small}

This is a long-tail distribution, reflecting the increasing difficulty of maintaining correctness over longer sequences.

This metric correlates well with user adoption rates and better reflects the model's actual capabilities in code completion. Moreover, the probability distribution of this metric aligns with the correctness distribution of the model's output, making it compatible with loss functions and providing clear guidance for model training.
Unlike LCS-based metrics that allow non-contiguous matches, our definition emphasizes continuity starting from the first character — a key aspect of interactive code writing and editing.

\noindent\textbf{ROUGE-LCP}: Inspired by the normalization concept of ROUGE-L, we further propose a normalized evaluation metric based on LCP, defined as follows:
\begin{small}
\begin{equation*}
\text{ROUGE-LCP}(S, R) = \frac{\text{LCP}(S, R)}{|R|}
\end{equation*}
\end{small}

This metric divides the LCP value by the length of the reference sequence $|R|$, enabling fair comparisons across samples of different lengths. Since LCP strongly correlates with user adoption behavior, we have reason to believe that ROUGE-LCP can also effectively reflect the practicality of model outputs in interactive scenarios. Experimental results also validate this hypothesis.
To further understand the statistical properties of ROUGE-LCP and establish a mathematical connection with Exact Match (EM), we introduce an auxiliary parameter $|S_{\text{ext}}|$, representing the length of the portion of the model's output that extends beyond the reference sequence when $\text{LCP}(S, R) = |R|$.
Based on this, we can model the probability distribution of ROUGE-LCP as a mixed distribution, as follows:

\begin{small}
\begin{equation*}
\begin{aligned}
  \MoveEqLeft P(\text{Rouge-LCP}(S, R)) = \\
  &
  \begin{cases}
    P\left( \dfrac{\text{LCP}(S, R)}{|R|} \right), \hspace{0.5em} \text{if } \text{LCP}(R,S) < |R|; \\
    P(\text{EM}(S, R)) , \hspace{0.5em} \text{if } \text{LCP}(R,S) = |R|\hspace{0.2em} and \hspace{0.2em} S=R; \\
    \\ 
    P\left( \dfrac{\text{LCP}(S, R) + |S_{\text{ext}}|}{|R|} \right), \\
    \hspace{4em} \text{if } \text{LCP}(R,S) = |R|\hspace{0.2em} and \hspace{0.2em} S \neq R;
  \end{cases}
\end{aligned}
\end{equation*}
\end{small}

According to the Central Limit Theorem, the length of the reference text $|R|$ typically follows an approximately Gaussian distribution in real-world settings.
This formula reveals the dual nature of ROUGE-LCP:

When the output prefix does not fully match the reference text, the metric value is less than 1, corresponding to partially correct cases;
When the output prefix fully matches the reference text, the metric value is 1, and the output may continue generating additional content (i.e., $|S_{\text{ext}}| > 0$), which should also be included in the evaluation scope.

Thus, the distribution of ROUGE-LCP can be viewed as the superposition of two sub-distributions:

\noindent\textbf{Partial Match Component}: Captures the model's ability to generate partially correct prefixes;

\noindent\textbf{Exact Match Component}: Measures the model's ability to not only generate correct prefixes but also continue producing useful content beyond that.

This modeling approach allows us to analyze the model's performance at different stages within a unified framework and provides a theoretical basis for further exploring its relationship with user adoption behavior. In subsequent experiments, we will separately analyze the correlation between these two components and user adoption rates.

\subsection{Repository-Level Code Corpus Processing Framework}

In this section, we describe in detail the corpus preprocessing pipeline, AST-based semantic segmentation, and the construction process of the code knowledge graph. Corpus preprocessing aims to improve the quality of training corpora by reducing noise and irrelevant information; AST-based semantic segmentation ensures the integrity of semantic units and prevents semantic disruption; while the code knowledge graph focuses on further constructing and enhancing the relationships between semantic units, thereby improving the model's global and cross-library semantic understanding capabilities.

To ensure that the generated knowledge graph corpora are of high quality, we have established a systematic preprocessing workflow, including key steps such as data filtering, data cleaning, and deduplication. Specific filtering rules, cleaning methods, and deduplication processes are detailed in Appx.\ref{sec:appendix_pipeline}.

\subsubsection{Syntax-Aware Semantic Unit Extraction via AST}
To enhance the structural awareness of pretraining corpora, we propose an AST-based semantic segmentation method as an alternative to traditional random or sliding-window masking strategies based on tokens. This method uses semantically closed subtrees in the AST as segmentation units, ensuring the structural integrity and contextual continuity of the masked units. Specifically, the method includes the following four steps: First, tools like Tree-sitter are used to parse the source code and extract semantically complete AST subtrees, such as function bodies and conditional branches. Second, a subtree is randomly sampled as the masking target, replaced with a placeholder, and concatenated with its preceding and succeeding contexts to form the training input. Third, structural integrity checks are performed to ensure that the masking operation does not disrupt the syntactic parsability of the remaining code. Finally, a granularity control parameter \(\theta\) is introduced to adjust the size of the subtrees, supporting multi-scale structural modeling. This method can be completed in linear complexity and is suitable for large-scale code corpus construction. For the formal modeling, algorithm, and comparative analysis with greedy segmentation, please refer to  Appx.\ref{sec:appendix_ast}.


\subsubsection{Structure-Preserving and Semantically-Reordered Code Graph}
After AST-based semantic unit extraction, segmentation, and completeness verification, we build a Structure-Preserving and Semantically-Reordered Code Graph (SPSR-Graph) to generate training corpora that maintain global call consistency. The construction proceeds in two stages: (1) semantic-unit extraction, where we parse the vertical-domain codebase to obtain self-contained units such as functions, structs, and classes; and (2) semantic-relationship graphing, where we connect these units with directed edges that encode calls, references, and inclusions. Traversing this graph along call paths allows us to reorder source code into contextually aligned sequences, enriching the structural depth of the corpus and enabling repository-level, cross-library context modeling.

To further extend the structural depth of training corpora and enhance cross-library context modeling, we propose organizing semantic units into structured graphs, constructing a semantic dependency graph named SPSR-Graph. The graph construction process is divided into two stages: semantic unit extraction and semantic relationship graphing.
The SPSR-Graph construction process consists of two stages:

The first stage is element extraction. We use an AST parser to parse the entire codebase and extract all top-level semantic units \(\nu_i \in \boldsymbol{\nu}\), such as function bodies, structs, and class definitions. Each \(\nu_i\) is semantically complete and stored in a structured database for subsequent calls.

The second stage is relationship extraction and graph construction. We represent the code graph as \(\boldsymbol{\Gamma} = (\mathcal{V}, \epsilon)\), where $\mathcal{V}$ is the set of nodes, i.e., the extracted semantic units; $\epsilon \subseteq \mathcal{V} \times \mathcal{V}$ is the set of directed edges. If there is a call relationship \(\nu_i \rightarrow \nu_j\), we define the edge \((\nu_i, \nu_j) \in \epsilon\). Edge types can support, but are not limited to: direct function calls (Direct Call), member references (Member Reference), type dependencies (Type Usage), macro or template expansions (Macro Expansion), and file inclusions (Include Dependency). To preserve contextual integrity, graph construction supports node attribute enhancement (e.g., definition location, module affiliation, syntax type labels) and edge type annotations, further enhancing the graph's semantic capacity.


On the directed graph \(\boldsymbol{\Gamma}\), we use directed BFS to search for all paths \(\mathcal{P}\) with depth \(d \leq D\):
\begin{small}
\begin{equation*}
\label{eq:semantic-paths1}
\begin{split}
\mathcal{P} = \{p_k = (\nu_{k_1}, \nu_{k_2}, ..., \nu_{k_m}) \mid \nu_{k_i} \in \mathcal{V},\ m \leq D\}
\end{split}
\end{equation*}
\end{small}
Path selection supports multiple strategies: forward call expansion (Forward Call Expansion), field access chain expansion (Field Access Expansion), and header inclusion prioritization (Header Inclusion). Each path \(p_k\) is mapped to the following training sample:
\begin{small}
\begin{equation*}
\label{eq:semantic-paths2}
\begin{split}
\text{Sample}(p_k) = \nu_{k_1} \oplus \nu_{k_2} \oplus \ldots \oplus \nu_{k_m}
\end{split}
\end{equation*}
\end{small}
where \(\oplus\) denotes structure-aware concatenation.

To enhance the model's cross-file structural modeling capability, we insert file path information and structural comments during concatenation:
\begin{small}
\begin{equation*}
\label{eq:semantic-paths3}
\begin{split}
\nu_{k_i} \mapsto \texttt{/* file: path/to/file */} \oplus \text{code}
\end{split}
\end{equation*}
\end{small}
The following algorithm explicitly shows the training sample construction process for SPSR-Graph. First, the system constructs the graph structure using the extracted AST semantic units, where each node represents a semantically complete code unit, and edges represent call or reference relationships. Then, breadth-first traversal (BFS) is used to enumerate all semantic paths with depth not exceeding $D$.

During traversal, for each valid path in the graph, the algorithm sequentially loads the source code fragments corresponding to each node in the path and embeds structural annotation information at cross-file boundaries. Finally, the structured fragments contained in the entire path are concatenated in dependency order to form a training sample with global semantic consistency, which is stored for subsequent language model pretraining. Let the total number of nodes be \(n\), the average outdegree be \(d\), and the maximum path depth be \(D\). Then, the complexity is: $\mathcal{O}(n + nd + n \cdot d^D \cdot m)$.

This process preserves both syntactic structural integrity and contextual consistency while achieving corpus reordering along call paths, enabling the model to explicitly encounter and model cross-function and cross-module structural dependencies during training.

\SetKw{KwEach}{for each}
\SetKwFunction{BFSPaths}{BFSPaths}
\SetKwFunction{Tag}{Tag}
\SetKwProg{Fn}{Function}{}{end}

\begin{algorithm}[ht]
\caption{SPSR-Graph Generator}
\small
\KwIn{AST‐unit DB $\mathtt{ASTDB}$; max depth $D$}
\KwOut{structure–aware sample set $\mathcal{S}$}
$\Gamma \gets \textsc{InitGraph}()$
\KwEach{$u \in \mathtt{ASTDB}$}{ $\Gamma.\textnormal{addNode}(u)$ }
\KwEach{$(u,v)\in \textsc{CallPairs}(\mathtt{ASTDB})$}{ $\Gamma.\textnormal{addEdge}(u\!\rightarrow\!v)$ }
$\mathcal{S}\gets\emptyset$\;
\KwEach{$p\in\BFSPaths(\Gamma,D)$}{
    $s\gets\textsc{Concat}(\,\Tag(u)\mid u\in p\,)$\;
    $\mathcal{S}\gets\mathcal{S}\cup\{s\}$\;
}
\Return $\mathcal{S}$
\BlankLine
\Fn{\BFSPaths{$\Gamma, D$}}{%
  \KwRet{all directed paths of $\Gamma$ with length $\le D$ (BFS)}}

\Fn{\Tag{$u$}}{%
  \uIf{\textsc{CrossFile}$(u)$}{\KwRet{\texttt{/* file: }$u.$file\texttt{ */}$\;\Vert$ code($u$)}}
  \Else{\KwRet{code($u$)}}}
\end{algorithm}

\section{Experiments}
\subsection{Experimental Setup}
This experiment uses Qwen2.5-7B-Coder as the base model \cite{hui2024qwen25codertechnicalreport}, adopting a 28-layer Transformer architecture with a total of approximately 7.6 billion parameters.
The model is pretrained on a 0.6B C/C++ code corpus from the communication domain and fine-tuned on approximately 60,000 lines of code corpus from the same domain. The fine-tuning corpus includes fields such as context, the line to be completed, and similar code retrieved via RAG.
Model training was conducted on a single server equipped with 8 NVIDIA A100 80GB GPUs.
We evaluated model performance using EM, LCP, and BLEU on user data.

We collected on-the-fly completion data from March 3, 2025, to April 24, 2025, using data logging for evaluation and analysis. The data source is ZTE-Code-Copilot.
The logged information includes timestamps, trigger points, programming languages, model predictions, context, and reference answers (i.e., the content confirmed by the user pressing the Enter key).

During the data preprocessing stage, we filtered out duplicate samples and data where the context contradicted the reference answers, resulting in a final dataset of 10,769 valid entries. Among these, 6,100 entries were from March 3 to March 31, and 4,669 entries were from April 1 to April 24. We divided this data into three groups for analysis: the first group covers March 3 to March 31; the second group covers April 1 to April 24; the third group spans the entire period.

\subsection{LCP and ROUGE-LCP Distribution and Correlation with Adoption Rate}
We first analyze the distribution of LCP and its correlation with the adoption rate in the three datasets.

\begin{figure}[t]
  \centering
  \includegraphics[width=1\columnwidth]{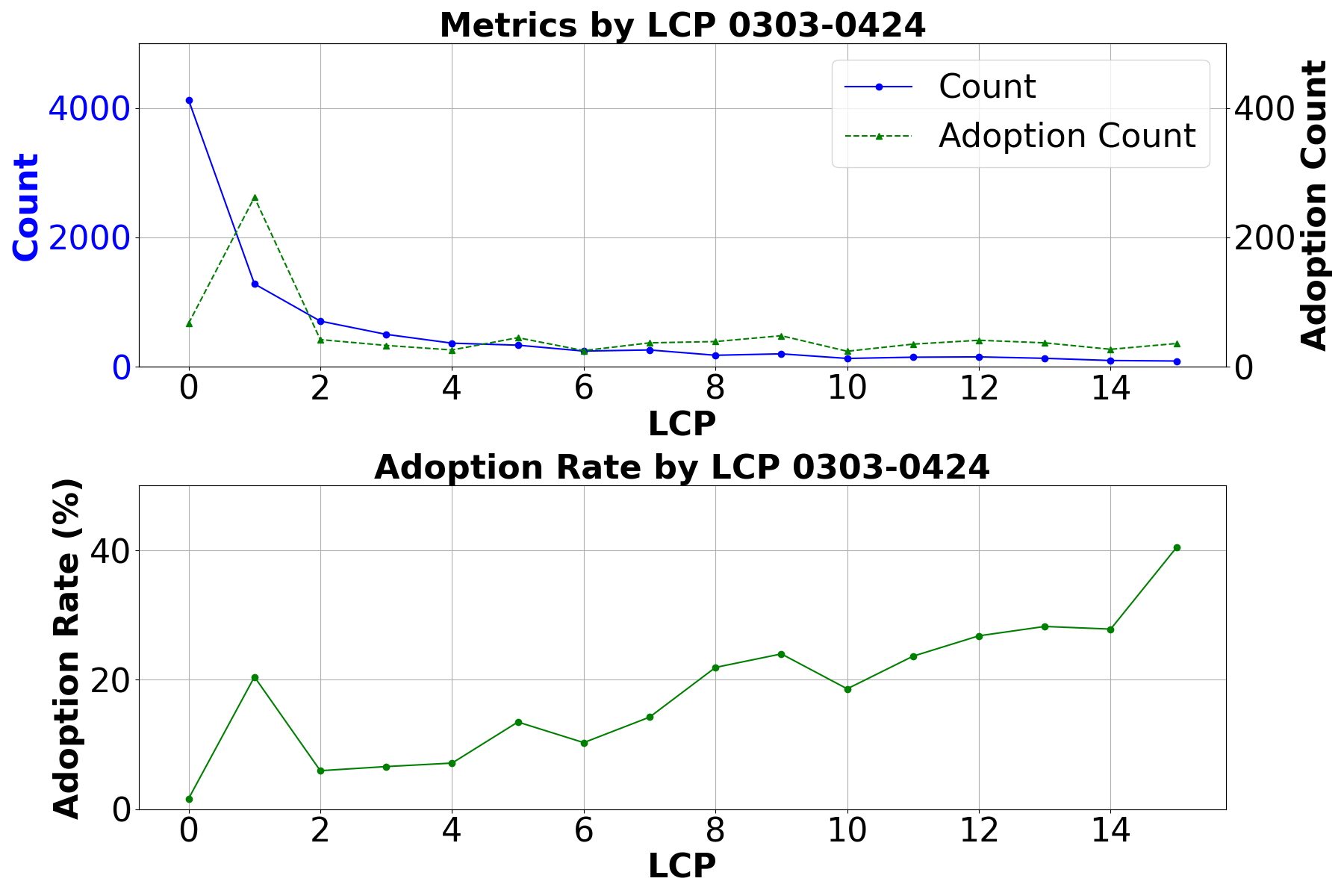}
  \caption{LCP Distribution and Its Relationship with Adoption Count and Adoption Rate}
  \label{fig:LCP Distribution and Its Relationship with Adoption Count and Adoption Rate}
\end{figure}

As shown in \Cref{fig:LCP Distribution and Its Relationship with Adoption Count and Adoption Rate}, the distribution of LCP has a clear long-tail property, which aligns well with the probabilistic structure derived in our theoretical modeling. Furthermore, we calculated the Pearson correlation coefficient (Pearson's r) between LCP and the adoption rate and found a significant positive correlation. As shown in \Cref{tab:R AND P}, across all time periods, the r values were above 0.6, and the P values were below 0.05, fully validating the effectiveness of the LCP metric.

Additionally, when LCP = 1, the adoption rate shows a local peak. This phenomenon can be attributed to users often adopting punctuation marks or syntactic structures such as ",",":",";" during code completion. Although these elements carry low information content, they still contribute to improving coding efficiency.

Next, we analyze the relationship between ROUGE-LCP counts, adoption counts, and adoption rates. Specifically, when $\text{LCP}(S, R) = |R|$, we split this scenario into two parts: the exact match (EM = 1) component and the partial extension component represented by $\frac{\text{LCP}(S, R)+|S_{\text{ext}}|}{|R|}$, evaluating their respective relationships with user adoption rates.

\begin{figure}[t]
  \centering
  \includegraphics[width=1\columnwidth]{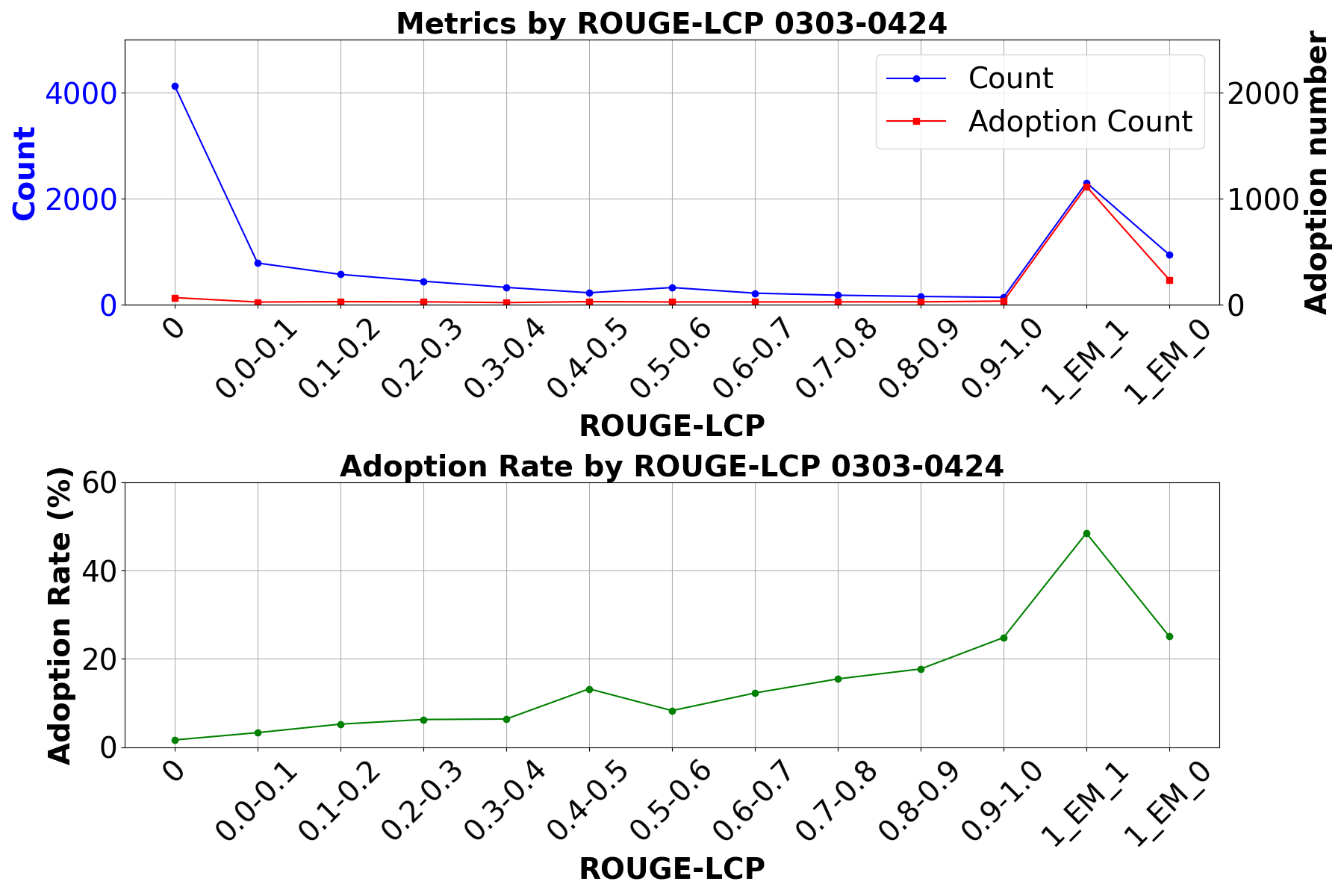}
  \caption{ROUGE-LCP Distribution and Its Relationship with Adoption Count and Adoption Rate}
  \label{fig:ROUGE-LCP Distribution and Its Relationship with Adoption Count and Adoption Rate}
\end{figure}

As shown in \Cref{fig:ROUGE-LCP Distribution and Its Relationship with Adoption Count and Adoption Rate}, the results indicate that the distribution of ROUGE-LCP follows a mixed distribution, consistent with our expectations. As the ROUGE-LCP value increases, the model’s adoption rate generally rises, peaking at EM = 1. Even when the model output exceeds the reference sequence length (i.e., $|S_{\text{ext}}| > 0$), there is still a certain level of adoption rate, suggesting that users tend to adopt and modify partially correct outputs. More results can be found in Appx.\ref{sec:appendix_CORRE}

\begin{table}[!ht]
    \centering
    \scalebox{0.8}{ 
        \begin{tabular}{llll}
        \hline
            \textbf{Metric} & \textbf{0303-0331} & \textbf{0401-0424} & \textbf{0303-0424} \\ \hline
            \textbf{r} & 0.9107 & 0.6952 & 0.8707  \\ 
            \textbf{P} & 0.0000 & 0.0028 & 0.0000 \\ \hline
        \end{tabular}
    }
    \caption{Pearson Correlation Analysis Between LCP and Adoption Rate}
    \label{tab:R AND P}
\end{table}


\subsection{Comparison with General Metrics}
To improve statistical stability, we filtered out data points with fewer than 100 completions per day, as these samples had low adoption counts, which could lead to significant fluctuations in adoption rates. Subsequently, we calculated the correlations between LCP, LCS, ROUGE-LCP, ROUGE-L, EM, and adoption rate on a daily basis to verify the effectiveness of the proposed metrics in reflecting user perception.

\begin{figure*}[t]
  \centering
  \includegraphics[width=0.95\linewidth]{ 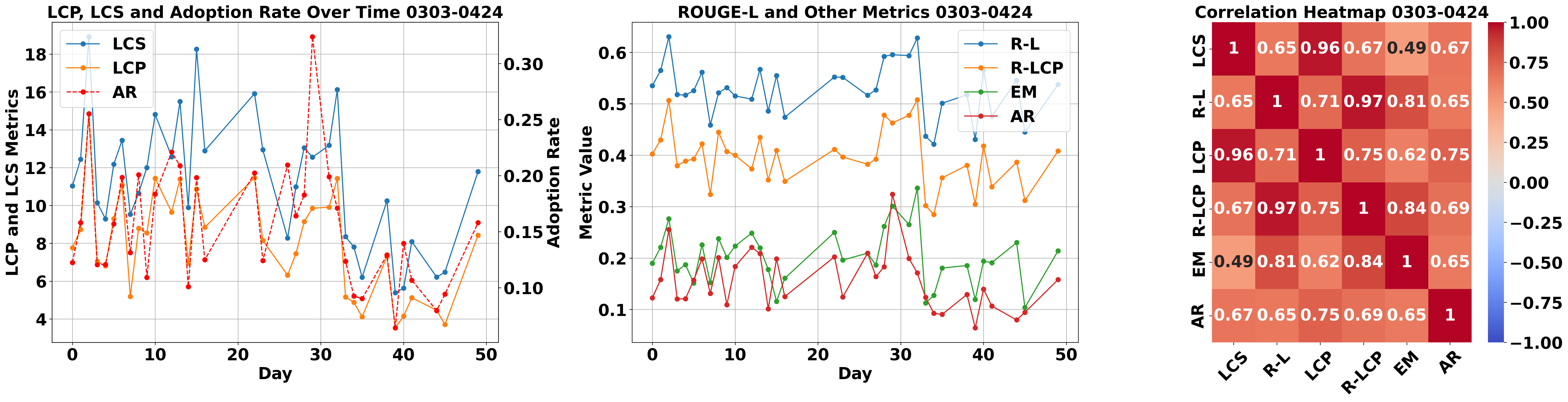}
  \caption{Daily Metric and Adoption Rate Distributions, Heatmap of Correlation Between Evaluation Metrics and Adoption Rate. R-L refers to ROUGE-L, R-LCP refers to ROUGE-LCP, and AR refers to Adoption Rate.}
  \label{fig:Heatmap}
\end{figure*}

As shown in \Cref{fig:Heatmap} and \Cref{tab: LCP and adopt_rate}, by observing the correlation heatmap, we found that compared to commonly used code completion metrics, LCP showed the strongest correlation with the adoption rate, with r values generally exceeding 0.7 and P values below 0.05; ROUGE-LCP followed closely. These results further validate that the proposed LCP and ROUGE-LCP metrics are better at capturing user intent and adoption behavior than general-purpose metrics. More results can be found in Appx.\ref{sec:appendix_LCP}
\begin{table}[!ht]
    \centering
    \scalebox{0.72}{
    \begin{tabular}{lllll}
    \hline
    \textbf{Metric}    & \textbf{0303-0331} & \textbf{0401-0424} & \textbf{0303-0424} & \textbf{}        \\ \hline
    \textbf{}          & \multicolumn{3}{c}{r}                                        & P\textless{}0.05 \\
    \textbf{LCP}       & \textbf{0.7145}    & \textbf{0.7862}    & \textbf{0.7485}    & \checkmark                \\
    \textbf{ROUGE-LCP} & 0.6224      & 0.7186       & 0.6867       & \checkmark                \\
    \textbf{LCS}       & 0.5863             & 0.7184             & 0.6677             & \checkmark                \\
    \textbf{ROUGE-L}   & 0.5729             & 0.6914             & 0.6498             & \checkmark                \\
    \textbf{EM}        & 0.5917             & 0.7037             & 0.6464             & \checkmark                \\ \hline
    \end{tabular}
    }
    \caption{Pearson Correlation Analysis Between Common Evaluation Metrics (LCS, LCP, ROUGE-L, ROUGE-LCP, EM) and Adoption Rate}
    \label{tab: LCP and adopt_rate}
\end{table}

\subsection{Impact of Different Pretraining Corpus Strategies}

This section sets up four groups of pretraining corpora for model pretraining, followed by fine-tuning using the same fine-tuning corpus described above, aiming to evaluate the impact of the proposed structure-aware corpus processing strategy on the "on-the-fly completion" task. 

\noindent\textbf{Pipeline}: Applying a data filtering, cleaning, and deduplication pipeline to the raw corpus;

\noindent\textbf{AST}: Applying AST-based semantic segmentation on top of the pipeline-processed corpus;

\noindent\textbf{\boldmath{$KG_F$}}: Constructing function-level code graph corpora based on AST-based semantic segmentation;

\noindent\textbf{\boldmath{$KG_{FS}$}}: Further introducing struct-level code graph corpora.

\begin{table}[!ht]
    \centering
    \scalebox{0.8}{
        \begin{tabular}{lcccccc}
        \toprule
        \textbf{} & \multicolumn{3}{c}{\textbf{C++}} & \multicolumn{3}{c}{\textbf{C}} \\
        \cmidrule(lr){2-4} \cmidrule(lr){5-7}
        \textbf{Corpus} & \textbf{EM(\%)} & \textbf{LCP} & \textbf{Blue} & \textbf{EM(\%)} & \textbf{LCP} & \textbf{Blue} \\
        \midrule
        Pipeline & 16.54 & 5.2 & 29.33 & 17.31 & 5.1 & 29.21 \\
        + AST & 16.72 & 4.8 & 27.83 & 17.70 & 5.1 & 28.90 \\
        + $KG_F$ & 17.84 & 5.2 & 29.47 & 20.36 & \textbf{5.8} & 31.64 \\
        + $KG_{FS}$ & \textbf{18.55} & \textbf{5.2} & \textbf{29.73} & \textbf{20.51} & 5.7 & \textbf{31.79} \\
        \bottomrule
        \end{tabular}
    }
    \caption{Performance comparison of different pretraining corpus strategies on the "on-the-fly completion" task}
    \label{overall}
\end{table}

As shown in Table~\ref{overall}, first, replacing random FIM with semantically complete AST-based segmentation further improves EM, demonstrating the benefit of consistent semantic unit boundaries in enhancing the model's perception of completion scope. Although BLEU slightly decreases, LCP remains stable, indicating that structural segmentation has no negative impact on prefix prediction.

Furthermore, the function-level code knowledge graph brings significant improvements, especially in C, where EM increases by 2.66\% and BLEU by 2.74 compared to AST alone. This suggests that after semantic reordering at the graph level, real cross-library information can be introduced, enabling the model to better capture global contextual relationships. Finally, adding struct-level knowledge graphs brings additional gains, particularly in complex languages like C++, where edge augmentation helps model the cross-node impact of type information.

In summary, the 3-stage evolution strategy—from data cleaning, structure-aware segmentation, to graph-based semantic reordering-improves the structural integrity and dependency consistency of the corpus, yielding consistent gains across multiple language tasks. The dual improvement in EM and BLEU, coupled with the steady growth in LCP, validates the practical effectiveness of structure-aware training samples.

\subsection{Impact of Code Knowledge Graph Breadth}

In this section, we explore the impact of graph traversal breadth on model performance. Since this task focuses on single-line or inline completion, which typically involves only local context, deep dependency modeling is unnecessary; thus, the traversal depth is fixed at 1. Considering the average dependency count of 3–5 in codebases and the model's maximum input length, we select $k={3,4,5,6,7}$ as the maximum breadth limit during graph construction.

\begin{figure}[ht]
\centering
\includegraphics[width=0.95\linewidth]{ 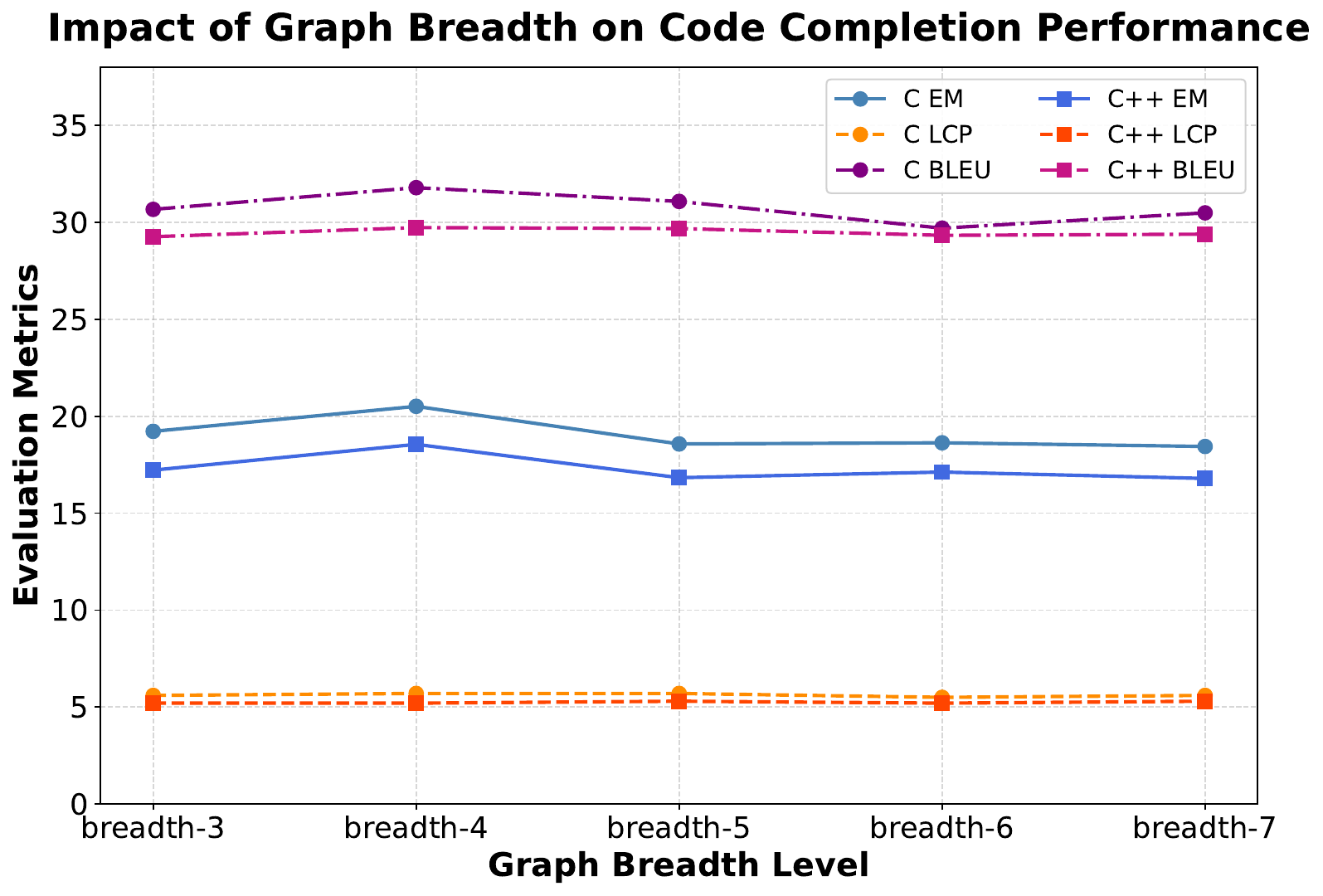}
\caption{Impact of code knowledge graph breadth on code completion performance}
\label{fig:kg-breadth}
\end{figure}

As shown in Figure \ref{fig:kg-breadth}, the breadth of the knowledge graph has a non-monotonic impact on model completion performance: when the breadth increases from 3 to 4, all metrics for both C and C++ improve significantly, reaching their peak; however, further increasing the breadth to 5–7 results in a slight decline or stabilization in performance. When the graph breadth becomes too large, it may introduce irrelevant context, distracting the model's attention and reducing completion accuracy;

Additionally, compared to C++, C shows greater sensitivity to changes in graph breadth, indicating that dependencies between functions in C are more concentrated, benefiting less from hints beyond the local scope. In contrast, C++'s more complex structure makes it less sensitive to edge expansion, reflecting its stronger adaptability to type references.

\section{Conclusion}
This paper proposes two evaluation metrics for the "on-the-fly completion" task: LCP and ROUGE-LCP. We theoretically analyze the relationship between these metrics and the model training objectives. Based on real user behavior data, we compare these metrics with other general evaluation metrics and validate the effectiveness of the proposed metrics in measuring user adoption rates. Additionally, we proposes a data processing method based on a Structure-Preserving and Semantically-Reordered Code Graph. Experimental validation demonstrates that this method significantly enhances the model's ability to perceive code structure and semantic dependencies, thereby effectively improving the model's performance in practical tasks.

\section{Limitations}
This paper focuses only on on-the-fly completion tasks and provides theoretical analysis and empirical validation for the proposed metrics. The current metrics in terms of model affinity are applicable only to autoregressive generative models, and their adaptability to other types of models remains to be further studied. The results indicate that when deploying large language models in real-world production environments, it is essential to design evaluation metrics tailored to the model's characteristics and business scenarios, in order to more accurately assess the model's practical performance and user experience.

\bibliography{main}
\appendix

\section{Corpus Preprocessing Details}
\label{sec:appendix_pipeline}

This appendix supplements the detailed implementation of the corpus preprocessing pipeline described in the methodology section of the main text, including data filtering, data cleaning, and deduplication.

\subsection{Data Filtering}

The data filtering stage aims to remove code samples that do not meet training standards. The specific rules include:

\textbf{Line Length Threshold Filtering}: Remove code samples where any line length falls outside the specified range to avoid abnormal or excessively short code.
    
\textbf{Average Line Length Filtering}: Remove samples with an average line length exceeding the threshold to ensure balanced code structure.
    
\textbf{Character Validity Ratio Filtering}: Discard samples with a low proportion of alphanumeric characters to exclude non-programmatic text.
    
\textbf{Total Character Count Filtering}: Delete files with a total character count below the minimum threshold to prevent fragmented code from entering the training set.
    
\textbf{File Type Filtering}: Explicitly exclude non-target language files (e.g., XML, HTML) to ensure the purity and specificity of the training corpus.

\subsection{Data Cleaning}

To improve corpus consistency and semantic validity, we introduce the following cleaning strategies:

\textbf{Code Formatting Standardization}: Use automated formatting tools to unify code style, enhancing consistency and parsability.
    
\textbf{Comment and Whitespace Cleanup}: Remove invalid comments, redundant spaces, and blank lines to minimize interference and avoid negative impacts on the model.

\subsection{Data Deduplication}

To prevent duplicate code samples from affecting training efficiency and model generalization, we adopt a two-stage deduplication strategy:

\textbf{Exact Deduplication}: Compute the SHA256 hash value for each code sample and remove completely identical samples.
    
\textbf{Fuzzy Deduplication}: Use the MinHash algorithm to detect structural and semantic similarity between samples, further removing highly redundant code fragments. This strategy not only reduces redundant data but also effectively preserves corpus diversity.

\section{Syntax-Aware Semantic Unit Extraction via AST}
\label{sec:appendix_ast}
This appendix supplements the complete implementation and theoretical analysis details of the AST-based structure-aware segmentation method described in the methodology section of the main text. Specifically, it includes: AST syntax modeling and semantic unit definition, pseudocode implementation of the AST semantic segmentation algorithm, and a theoretical analysis of traditional greedy segmentation.

Fill-in-the-Middle (FIM) is a widely used pretraining strategy for code completion tasks, where the core idea is to mask code segments and let the model predict the masked content. However, traditional FIM samples are often based on token-level greedy segmentation, such as fixed-step sequential partitioning or random cutting. While this approach is simple to implement, it can easily disrupt semantic structures in programs, such as functions, conditional branches, and loop bodies, thus limiting the model's ability to model structural information.
To address this, we propose an Abstract Syntax Tree (AST)-based semantic-aware segmentation method to enhance the consistency and integrity of structural units in training samples.

\subsection{AST Modeling and Semantic Unit Definition}

An abstract syntax tree is a structured representation of source code in terms of syntax. Formally, we represent an AST as an ordered tree $T = (V, E)$. Here, $\quad V = \{v_i\}_{i=1}^n$ is the set of nodes representing syntactic units such as keywords, expressions, and statement blocks; $E \subseteq V \times V$ is the set of edges, with each edge $(v_i, v_j)$ indicating that $v_j$ is a syntactic child node of $v_i$, forming a hierarchical structure recursively expanded from parent nodes based on grammar rules. Each node \(v_i\) has a label \(t_i\):
\[
\text{label}(v_i) = t_i, \quad t_i \in \mathcal{T}
\]
where \(\mathcal{T}\) is the set of non-terminal symbols defined by the language, such as \texttt{FunctionDefinition}.

Given a node \(v_i \in V\), we define the AST subtree rooted at \(v_i\) as \(T_s(v_i) \subset T\). This subtree contains all child nodes reachable from \(v_i\) via the set of syntactic edges \(E\). We refer to \(T_s(v_i)\) as a "semantically complete unit," as it forms a closed, independently modelable structural fragment in terms of both syntax and semantics, such as a function body or control flow block.

\subsection{Overview of AST Semantic Segmentation Method}

We propose a structure-aware training sample construction method based on abstract syntax trees, aiming to replace the traditional FIM sample generation strategy based on linear token segmentation. This method follows four key steps:

\begin{enumerate}
  \item \textbf{Semantic Unit Extraction}: Using syntax parsing tools like Tree-sitter, extract semantically closed AST subtrees from the source code. Each subtree \(T_s(v_i)\) represents a structurally modelable code unit (e.g., function definition, conditional branch, or loop body) with complete syntactic boundaries and internal semantic consistency.
  
  \item \textbf{Structure-Consistent Mask Construction}: Randomly sample a semantic unit \(S = T_s(v_i)\) from the candidate subtrees, constructing the input as:
  \[
  \text{Input} = P_1 ~ \texttt{<mask>} ~ P_2, \quad \text{Target} = S
  \]
  where \(P_1\) and \(P_2\) are the preceding and succeeding contexts of the subtree in the original source code, respectively. This design ensures that the generated training samples remain consistent with real-world code structures.
  
  \item \textbf{Semantic Completeness Verification}: After each masking operation, verify whether the remaining AST still constitutes a syntactically valid tree. If the masking disrupts the closure of the AST (e.g., breaking control statements or expression blocks), discard the candidate subtree to ensure the structural correctness of the training samples.
  
  \item \textbf{Structural Granularity Control Mechanism}: Introduce a masking granularity control parameter \(\theta\) to adjust the size of the masked subtree (e.g., number of tokens or nodes). By sampling \(\theta \in [\theta_{\min}, \theta_{\max}]\), this method supports multi-scale semantic modeling, ranging from micro-level expression masking to macro-level function masking.
\end{enumerate}
This method can be formalized as the following algorithm. We first parse the source code to construct the corresponding abstract syntax tree (AST), then traverse its structural semantic units. Under the constraints of granularity and structural completeness, we construct standard Fill-in-the-Middle training samples for each candidate subtree. The complete process is as follows:
\begin{algorithm}[ht]
\caption{AST Semantic FIM Cut}
\KwInput{Code file $C$; threshold $\theta$}
\KwOutput{A set of structure-aligned FIM training samples}

$AST \leftarrow \textnormal{Parse}(C)$\;
$Nodes \leftarrow \textnormal{ExtractSemanticSubtrees}(AST)$\;
$Masked \leftarrow \emptyset$\;

\KwEach{$T_s \in Nodes$}{
  \If{$\textnormal{Size}(T_s) \leq \theta$ \textbf{and} $\textnormal{IsComplete}(AST \setminus T_s)$}{
    $(P_1, P_2) \leftarrow \textnormal{ContextAround}(T_s)$\;
    $Sample \leftarrow \{ \textnormal{``input''}: P_1 + \langle\texttt{MASK}\rangle + P_2,\, \textnormal{``target''}: T_s \}$\;
    $\textnormal{Masked.append}(Sample)$\;
  }
}
\Return{$Masked$}
\end{algorithm}

We use depth-first traversal (DFS) to extract subtrees from the abstract syntax tree, with a time complexity of \(\mathcal{O}(|V|)\). The completeness check of the masked region can also be completed in linear time, with the same complexity of \(\mathcal{O}(|V|)\). Therefore, the overall segmentation process incurs linear overhead during the preprocessing stage, making it suitable for large-scale code pretraining tasks.

\subsection{Theoretical Analysis: Greedy Segmentation vs. Structural Disruption}

In traditional FIM task construction, samples are often generated using a greedy segmentation strategy, which divides the original code sequence into segments of fixed token window length \(\ell\). Assuming a function body contains \(L\) total tokens, it will be divided into \(k = \lceil L / \ell \rceil\) segments.

Since greedy segmentation does not perceive any syntactic structure, its segmentation boundaries may fall within semantic units such as function bodies, conditional statements, or loop bodies, causing semantic fragmentation. We define the structural preservation rate (the probability that a semantic unit remains unbroken) as:
\[
R_s = \frac{1}{k}
\]
while the remaining \(k - 1\) segments are structurally disrupted, with a proportion of \(1 - \frac{1}{k}\). As \(k\) increases, the structural disruption rate rises sharply, especially when dealing with large functions or deeply nested structures, significantly interfering with the model's ability to model program structures.

In contrast, AST-based semantic segmentation uses syntactic structures as boundaries, ensuring that segmentation units are complete semantic subtrees. This method theoretically guarantees that no semantic structure is broken, with a structural preservation rate of \(R_s = 1.0\). Thus, in structural modeling tasks, the AST segmentation strategy provides significantly stronger structural consistency guarantees.

\section{LCP and ROUGE-LCP Distribution and Correlation with Adoption Rate in all time periods}
\label{sec:appendix_LCP}
\begin{figure*}[t]
  \includegraphics[width=0.5\linewidth]{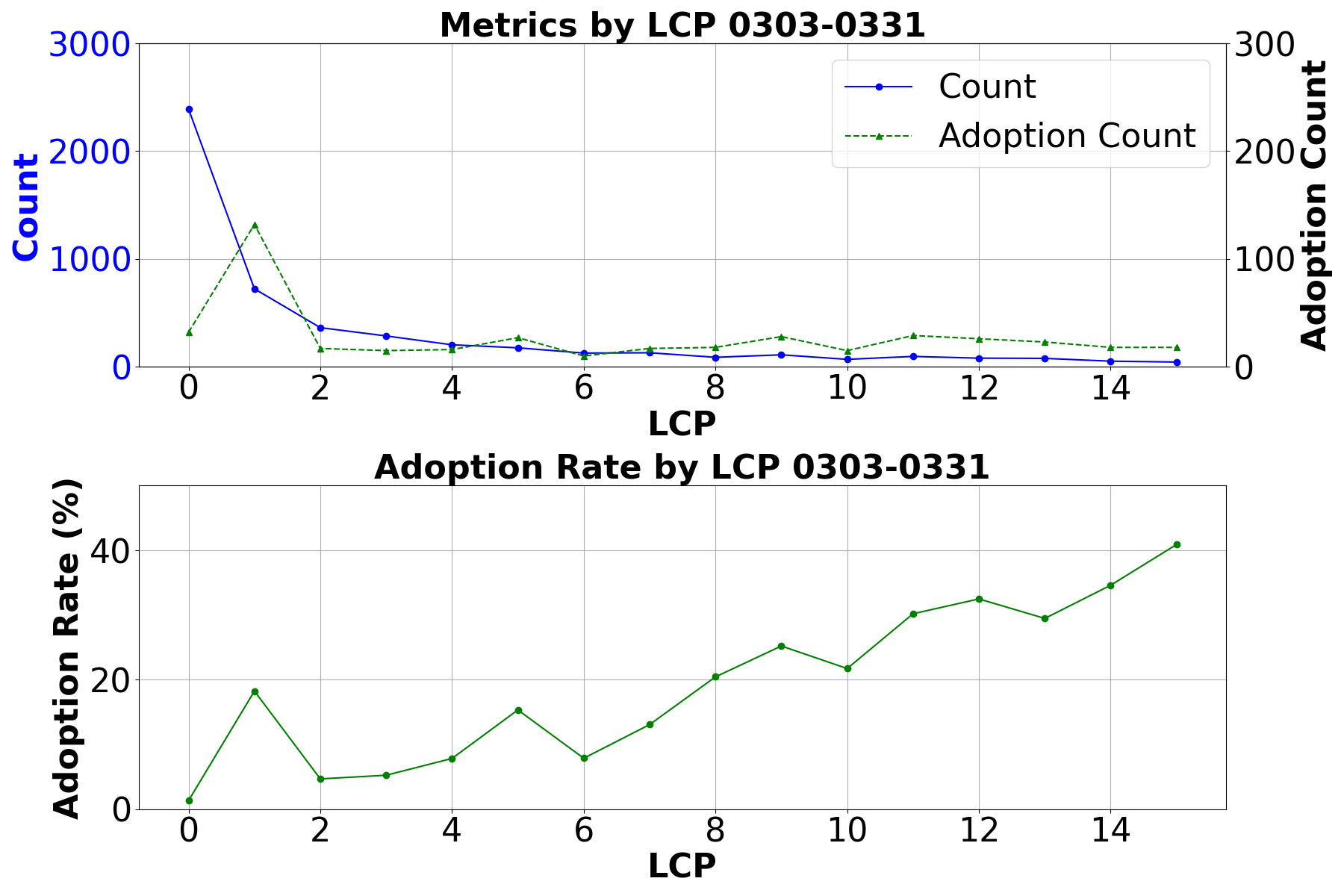} \hfill
  \includegraphics[width=0.5\linewidth]{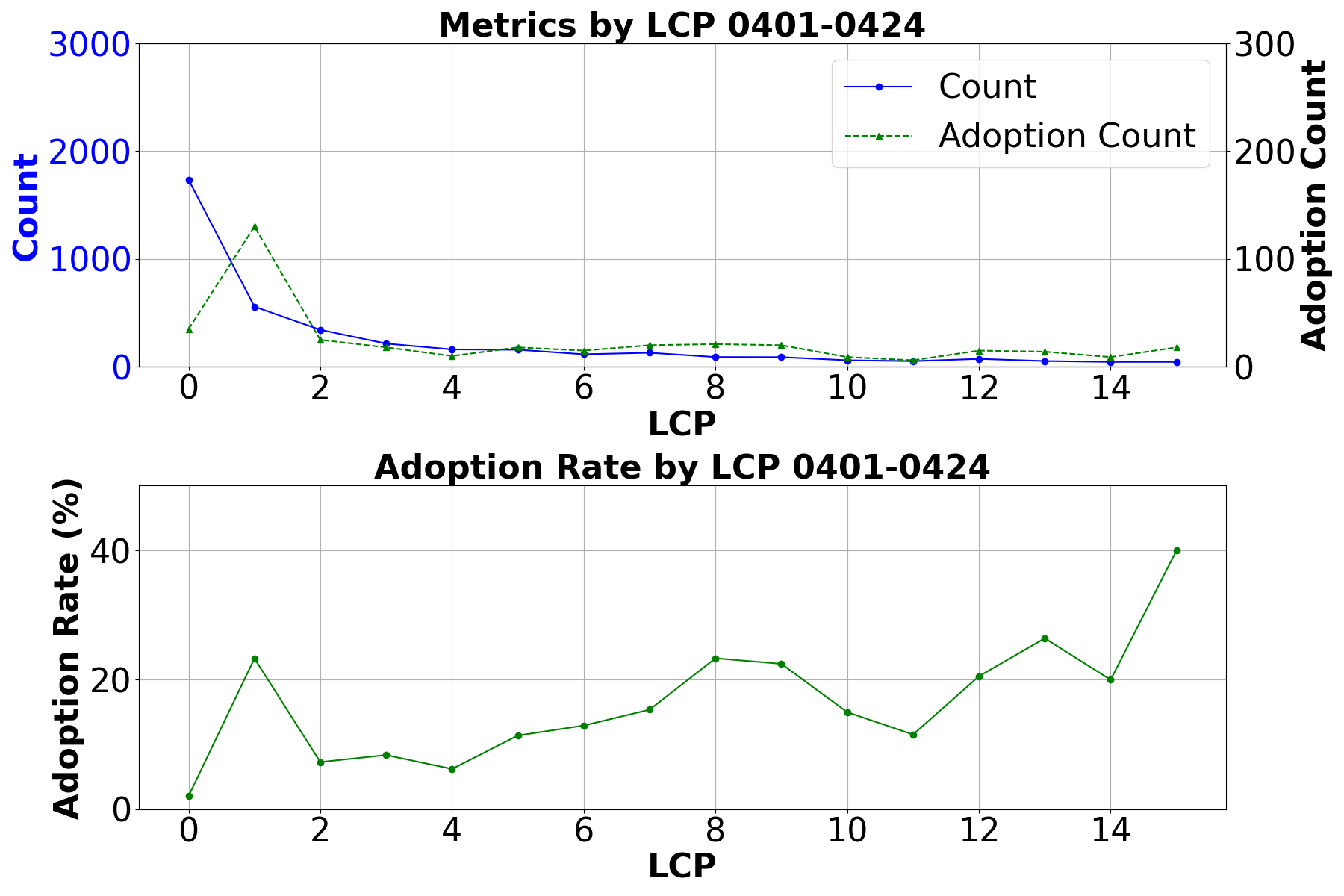} \hfill
  \caption {LCP Distribution and Its Relationship with Adoption Count and Adoption Rate}
 \label{fig:AP_1} 
\end{figure*}

\begin{figure*}[t]
  \includegraphics[width=0.5\linewidth]{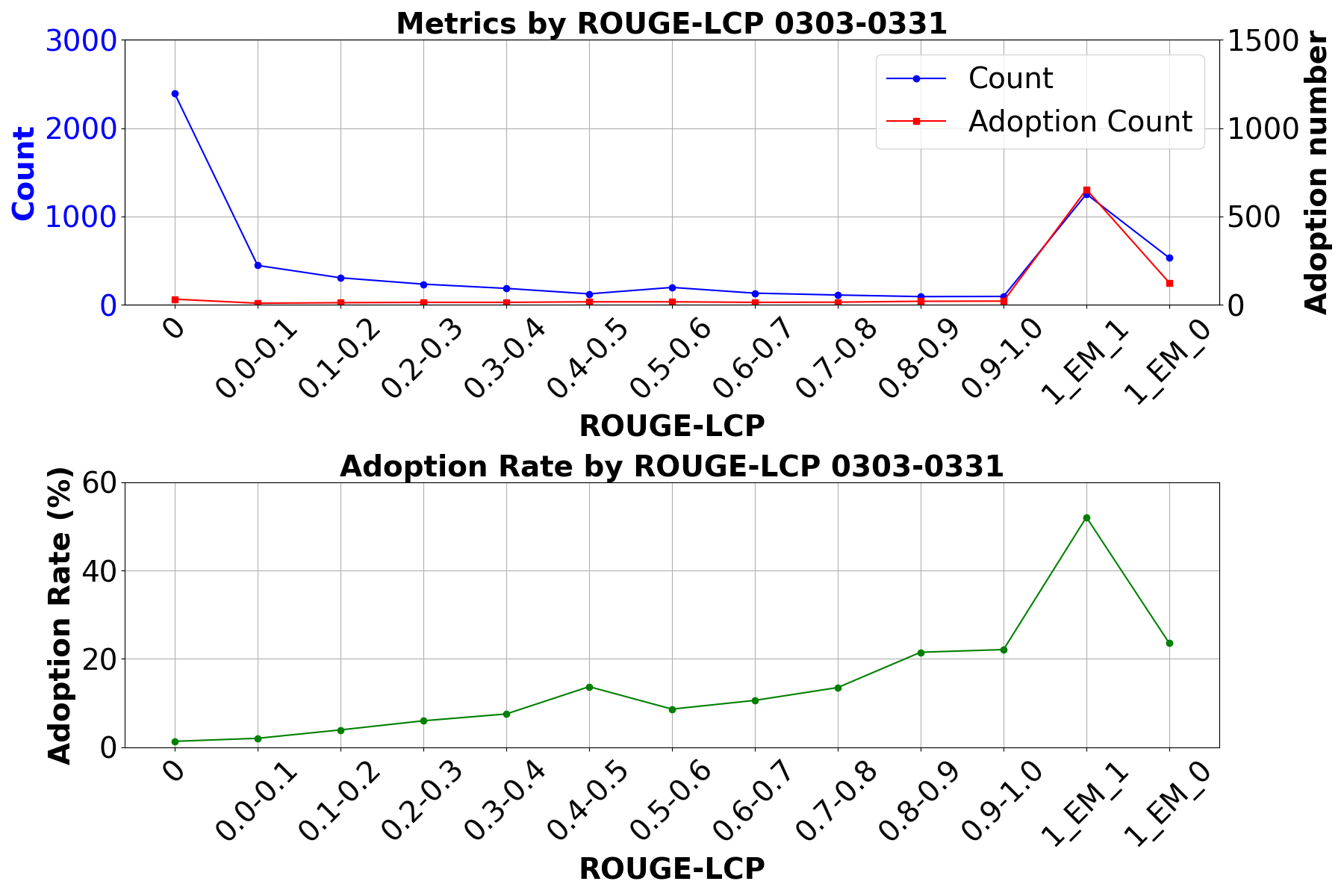} \hfill
  \includegraphics[width=0.5\linewidth]{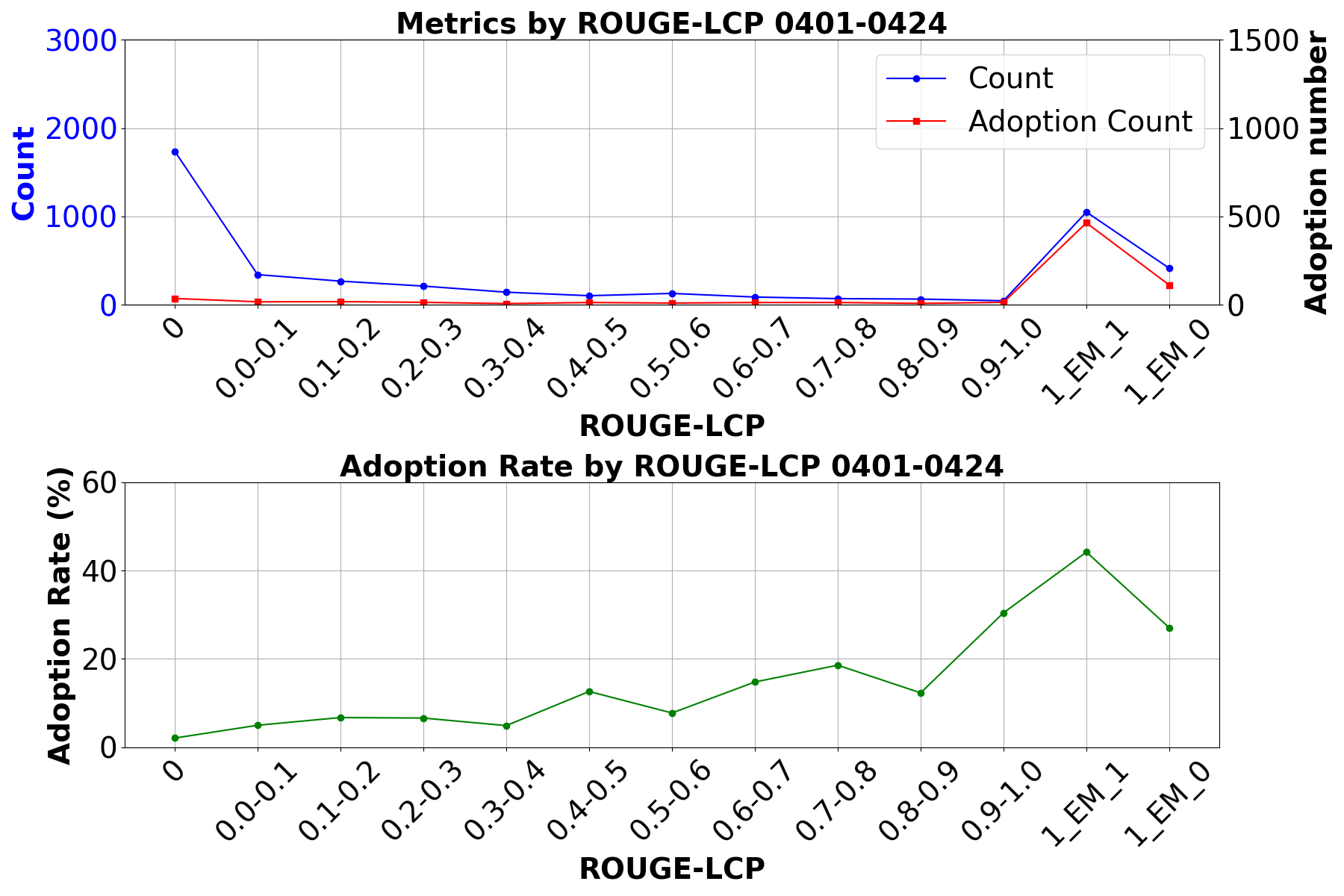} \hfill
   
  \caption {ROUGE-LCP Distribution and Its Relationship with Adoption Count and Adoption Rate}
  \label{fig:AP_2} 
\end{figure*}

Here, we present the LCP and ROUGE-LCP distributions and their correlation with the adoption rate for the periods from March 3 to March 31, April 1, and April 24. As shown in \Cref{fig:AP_1} and \Cref{fig:AP_2}. The three datasets consistently exhibit the same characteristics: the distribution of LCP shows a significant long-tail property, which aligns well with the probability structure derived in our theoretical modeling. From the figures, it can be observed that there is a significant positive correlation between LCP and the adoption rate. Additionally, when LCP equals 1, the adoption rates of all three datasets show a local peak. Next, we examine the relationship between the count, adoption frequency, and adoption rate of ROUGE-LCP across the three time periods. The results indicate that the distribution of ROUGE-LCP exhibits a mixed distribution, consistent with our expectations. As the ROUGE-LCP value increases, the model's adoption rate generally rises and reaches its highest point when EM = 1.

\section{LCP and ROUGE-LCP Comparison with General Metrics in all time periods}
\label{sec:appendix_CORRE}
\begin{figure*}[t]
  \includegraphics[width=1\linewidth]{ 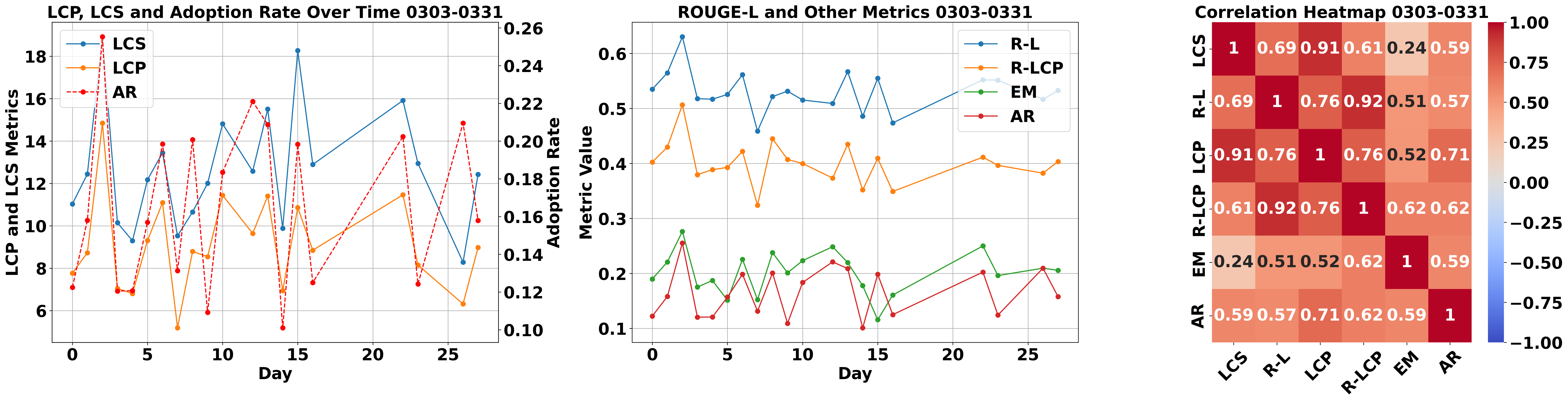} \\ 
  \includegraphics[width=1\linewidth]{ 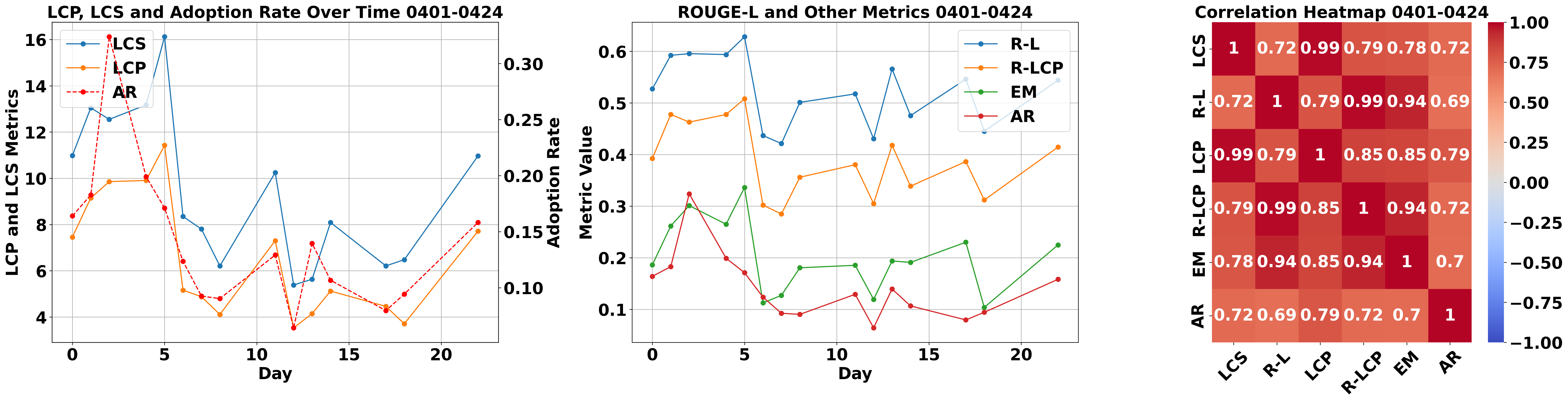} \\ 
  
  \caption{Daily Metric and Adoption Rate Distributions, Heatmap of Correlation Between Evaluation Metrics and Adoption Rate. R-L refers to ROUGE-L, R-LCP refers to ROUGE-LCP, and AR refers to Adoption Rate.} 
  \label{fig:AP_3} 
\end{figure*}

Here, we also present the Daily Metric and Adoption Rate Distributions, as well as the Heatmap of Correlation Between Evaluation Metrics and Adoption Rate for the periods from March 3 to March 31, April 1, and April 24. As shown in \Cref{fig:AP_3} We calculate the correlations between LCP, LCS, ROUGE-LCP, ROUGE-L, EM, and the adoption rate on a daily granularity to verify the effectiveness of the proposed metrics in reflecting user perception.

By observing the correlation heatmaps, we find that compared to commonly used code completion metrics, LCP shows the strongest correlation with the adoption rate across all time periods, with r values generally exceeding 0.7 and p-values all below 0.05; ROUGE-LCP ranks second. These results further validate that the proposed LCP and ROUGE-LCP metrics outperform general metrics in capturing user intent and adoption behavior.

\end{document}